\documentclass[12pt]{article}
\usepackage{amsmath,amsfonts,amssymb,bbm}
\usepackage{graphics}
\usepackage{amsfonts}
\usepackage{xcolor}

\usepackage[latin1]{inputenc}
\usepackage[pdftex]{graphicx}

\usepackage{lipsum}
\usepackage[color, leftbars]{changebar}

\usepackage{fullpage}
\usepackage{color}
\usepackage[american]{babel}
\usepackage{physics}

\makeatletter \@addtoreset{equation}{section}

\makeatletter\renewcommand\section{\@startsection {section}{1}{\z@}%
                                   {-3.5ex \@plus -1ex \@minus -.2ex}
                                   {2.3ex \@plus.2ex}%
                                   {\normalfont\large\bfseries}}
\renewcommand\subsection{\@startsection{subsection}{2}{\z@}%
                                     {-3.25ex\@plus -1ex \@minus -.2ex}%
                                     {1.5ex \@plus .2ex}%
                                     {\normalfont\bfseries}}

\newcommand{\be}{\begin{equation}}
\newcommand{\ee}{\end{equation}}
\newcommand{\bea}{\begin{eqnarray}}
\newcommand{\eea}{\end{eqnarray}}
\newcommand{\bmat}{\begin{bmatrix}}
\newcommand{\emat}{\end{bmatrix}}

\newcommand{\bbibitem}[1]{\bibitem{#1}\marginpar{#1}}

\def\Label#1{\label{#1}%
  \smash{\hbox to0pt{\raise1ex\hbox{\tiny[#1]}\hss}}}
\def\noLabels{\let\Label=\label}
\def\nobbibitem{\let\bbibitem=\bibitem}

\newcommand{\p}{\partial}

\usepackage{cite}

\begin{document}

\hspace*{0pt}\hfill  
\\

\vspace{1truecm}

\begin{center}
{\fontsize{21}{18} \bf Love Numbers for Extremal Kerr Black Hole}\\[14pt]
\end{center}

\vspace{.15truecm}

\begin{center}
{\fontsize{13}{18}\selectfont  Malcolm Perry $^{\rm a,b,c}$\footnote{\texttt{malcolm@damtp.cam.ac.uk}} and 
Maria J. Rodriguez $^{\rm d,e,f}$\footnote{\texttt{maria.rodriguez@gmail.com}}}
\end{center}
\vspace{.4truecm}

\begin{scriptsize}
 \centerline{{\it ${}^{\rm a}$ Department of Physics and Astronomy, Queen Mary University of London, 
 London E1 4NS, UK}}

  \vspace{.05cm}

 \centerline{{\it ${}^{\rm b}$ DAMTP, Centre for Mathematical Sciences, University of Cambridge, 
 Cambridge CB3 0WA, UK}}

  \vspace{.05cm}
  
   \centerline{{\it ${}^{\rm c}$ Trinity College, Cambridge, CB2 1TQ, UK}}

    \vspace{.05cm}

 \centerline{{\it ${}^{\rm d}$Department of Physics, Utah State University, Logan, UT 84322, USA}}

  \vspace{.05cm}

  \centerline{{\it ${}^{\rm e}$Black Hole Initiative, Harvard University, Cambridge, MA 02138, USA}}

  \vspace{.05cm}

 \centerline{{\it ${}^{\rm f}$Instituto de F\'\i sica Te\'orica UAM-CSIC,
 Madrid, 28049, Spain}}

\end{scriptsize}

 \vspace{.25cm}

\vspace{.3cm}
\begin{abstract}
\noindent

We perform a detailed study of the gravitational tidal Love numbers of extremal zero-temperature Kerr black holes. These coefficients are finite and exhibit the dissipative nature of these maximally spinning black holes. Upon considering the dynamical behavior of the tidal deformations of the extremal Kerr black holes, we provide explicit expressions of the Love numbers at low frequencies. Their calculation is simplified to specific formulas, which are directly derived using the Leaver-MST methods.


\end{abstract}

\newpage

\setcounter{tocdepth}{2}
\tableofcontents
\renewcommand*{\thefootnote}{\arabic{footnote}}
\setcounter{footnote}{0}

\section{Introduction}

Black holes are the most resilient compact objects in the cosmos. Classically, once a black hole forms, it can continue to grow by accreting matter or by merging with other black holes, increasing its mass and spin. However, under General Relativity (GR), black holes cannot be disrupted or die through disintegration. Even under drastic conditions, when placed in strong static external gravitational fields, black holes  maintain their gravitational dominance over spacetime, exhibiting zero tidal deformations \cite{damour2009black, binnington2009love, gurlebeck2015no, cardoso2017black, kol2012specul,Landry:2015zfa,Hui:2020xxx}. It was recently shown, that black holes may deform in response to time-varying (dynamical) tidal forces \cite{Charalambous:2021mea, Chia:2020yla, Saketh:2023bul,Perry:2023wmm}, but they cannot dissolve classically or lose their fundamental structure, the event horizon. 

Among them, extremal black holes, with their maximal spin and zero-temperature, are the most challenging realizations in GR. Unlike non-extremal black holes, extremal Kerr black holes were shown to amplify the effects of even minimal deviations from classical predictions \cite{horowitz2023extremal, Cano:2024bhh}.  More surprisingly, extremal black holes are believed to be abundant in the universe, as suggested by recent research \cite{Daly:2023axh,Daly_2019}. Observations indicate that these almost extremal black holes frequently occur at the centers of galaxies, where their extreme properties may influence the dynamics of their environments.

As observational evidence for gravitational waves becomes increasingly precise \cite{LIGOScientific:2016lio, maselli2017dynamical, flanagan2008tidal, Khanna:2016yow, wasserman2019lisa}, the study of tidal deformations of black holes in dynamic settings is becoming essential. Understanding how black holes respond to external tidal forces during binary mergers or other gravitational interactions is crucial for accurately modeling and interpreting the waveforms detected by gravitational wave observatories. These deformations can provide deeper insights into the properties of black holes and test fundamental aspects of GR and potential modifications in the strong-field regime.

\begin{figure}[h!]
  \centering
  \includegraphics[width=7.5cm]{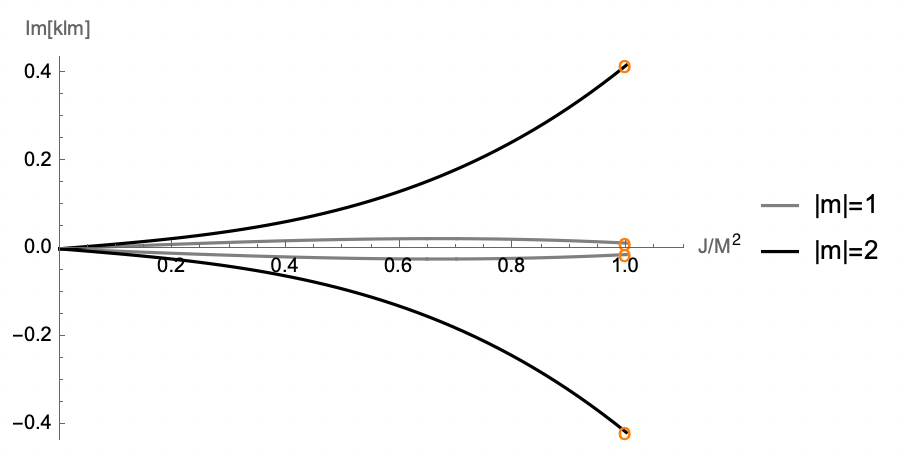}
  \includegraphics[width=7.5cm]{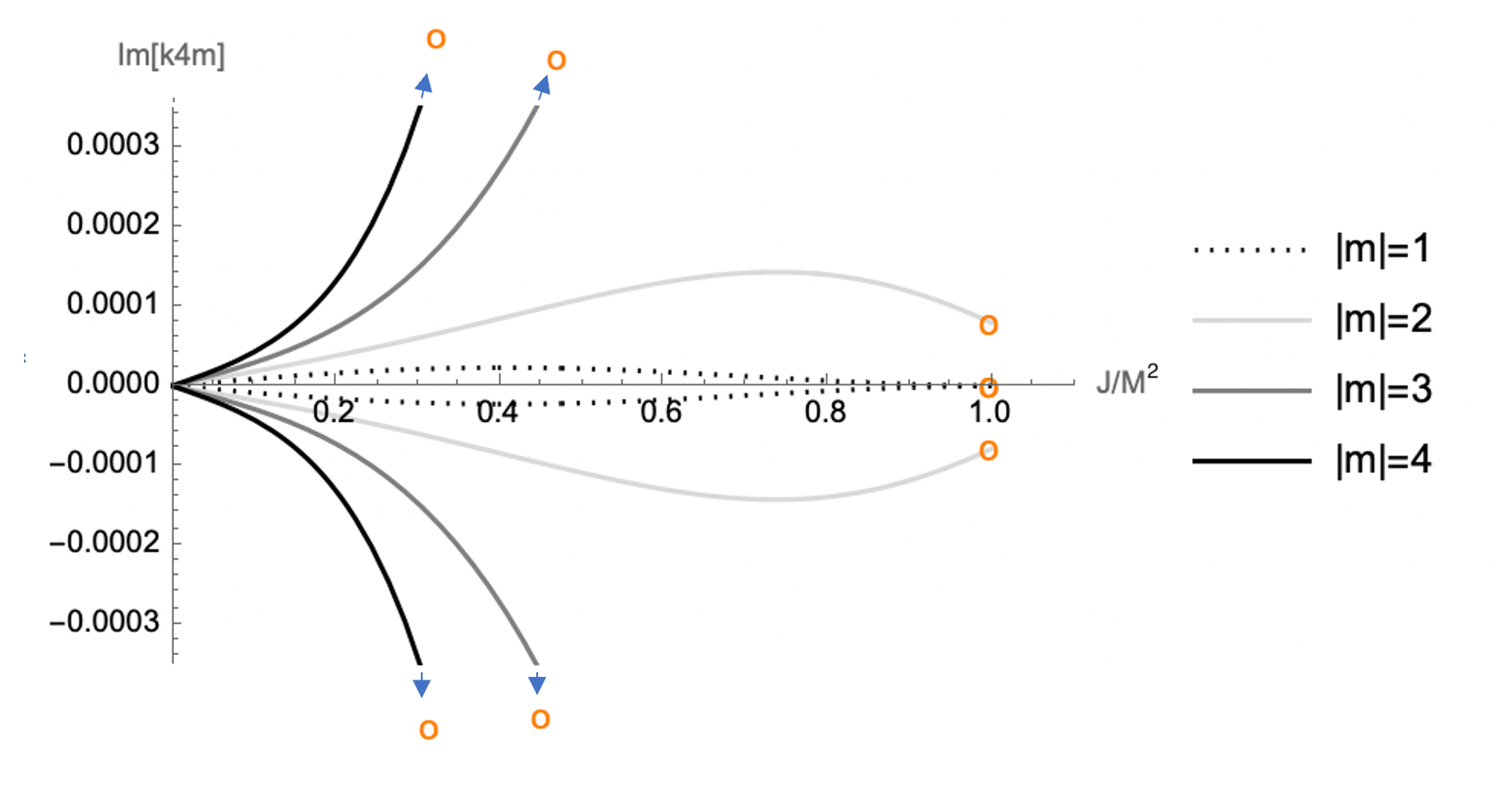}
  \caption{Static Love numbers of Kerr black holes $k_{\ell,m}$ for gravitational fields $s=2$ as a function of the angular momentum $J/M^2$. In this paper, we explicitly determine the endpoints of these curves, which correspond to black holes that are maximally rotating (extremal) and have zero temperature. We argue that the static Love numbers for extremal Kerr black holes are finite ({\it orange} points) except for $m=0$ e.g. taking values for example for $\ell=2$ respectively $k_{2,\pm m=0,1,2}=(0,\pm0.0130906,\pm0.418869)$ and for $\ell=4$ we find $k_{4,\pm m=0,1,2,3,4}=(0,\pm 1.545963 \times 10^{-7}, \pm 0.0000791533, \pm 0.00304292, \pm 0.0405265)$}
  \label{fig:love_number_l2}
\end{figure}

The rotation of a black hole has a significant effect on its tidal Love numbers, with the deformation coefficients showing varying behaviors based on the black hole's spin. To illustrate the relationship between spin in static Love numbers, we provide representations in Figure \ref{fig:love_number_l2}. This raises the intriguing question of whether extremal Kerr black holes at zero temperature display critical behavior similar to superconductors. Specifically, this paper explores whether extremal black holes dissipate energy at zero temperature and whether their (both static and dynamical) tidal dissipative coefficients vanish.

Note that as the black hole's (dimensionless) spin $J/M^2$ grows, the static tidal dissipative response increases. However, for every fixed values of the spheroidal eigenvalues $m, \ell$, a maximum is reached, after which the tidal deformability decreases, reaching a minimum fixed non-trivial value at extremality ($J/M^2=1$) where the black hole temperature vanishes. 

To date, analytical studies of tidal deformations for extremal Kerr black holes have made some progress in understanding these complex systems \cite{Bhatt:2024mvr,Bhatt:2024yyz,Kehagias:2024yzn}. In this paper, we utilize the formalism introduced by Casals et al. \cite{cardoso2017black} to systematically compute expressions for the Love numbers, which characterize the tidal deformations of extremal Kerr black holes. This approach allows for a clearer understanding of how extremal black holes respond to both static and dynamical tidal forces, resolving previous discrepancies and advancing the study of black hole tidal dynamics in strong gravitational fields.

The rest of the paper is organized as follows. In Section \ref{sec:intro} we introduce the Teukolsky equation relevant for deriving the tidal deformation coefficients of black holes. In Section \ref{sec:StaticEK} we solve the relevant equations to obtain the exact analytic expressions of the static Love number for zero temperature extremal Kerr black holes. We derive the dynamical tidal deformation coefficients for extremal Kerr black holes in Section \ref{sec:DynEK}, utilizing the techniques developed by Leaver, as well as Mano, Suzuki, and Takasugi (MST). In the low-frequency limit, we demonstrate that the static Love numbers become zero, while the dynamical dissipation coefficients remain non-zero. In the concluding section, we explore the significance of our findings and outline potential future directions.

\section{Basic Set-up}
\label{sec:intro}

The tidal Love numbers, quantifying how much the black hole deforms in response to the external tidal field, can be extracted from the asymptotic behavior of the gravitational field far from the black hole, where the perturbed metric can be compared to the tidal field.
The process begins by introducing a perturbation in the scalar field around the black hole:
\begin{eqnarray}\label{scalar}
\psi_s(r, \theta, \phi, t) =e^{-i  \omega t+i m\phi } \hat{R}_s(r) S_s(\theta) 
\end{eqnarray}
where $\hat{R}_s(r)$ is the radial function that depends on the radial coordinate $r$, $S_s(\theta)$ is the spin-weighted spheroidal harmonic that depends on the angular coordinate $\theta$, $\omega$ is the frequency of the perturbation, $m$ is the azimuthal quantum number, and $s$ is the spin weight of the perturbing field (e.g., $s=0$ for scalar fields, $s=\pm1$ for electromagnetic fields, and $s=\pm 2$ for gravitational waves). 
As famously demonstrated by Teukolsky (see, e.g., \cite{Teukolsky:1973ha,Teukolsky:1972my}), the governing differential equation for the scalar field \eqref{scalar} in an extremal Kerr background can be separated into radial and spheroidal angular components. This separation allows for the independent treatment of each equation.

For extremal Kerr black holes, in the standard Boyer-Lindquist parametrization of the Kerr metric, we set the mass \( M \) and spin parameter \( a \) to be proportional. The extremal condition is achieved when \( M = a \), which corresponds to the maximum allowed spin for a black hole of mass \( M \), where the horizon temperature approaches zero.
The radial part of the Teukolsky equation becomes
\bea\label{eq:radial}
&\!\!\!\!\!\!\!\!\!\!&\Bigg[\p_r\left( (r-M)^2 \p_r \right)+  \frac{M(m-2 M \omega)(m M+ 2 i (r-M)s+ 2 M (M-2 r)\,\omega)}{(r-M)^2}   \\
&\!\!\!\!\!\!\!\!\!\!& \quad\qquad\qquad\qquad\qquad\qquad + \omega^2 r^2 + 2(M\omega+is)\omega r + 2M \,\omega\,(2M\,\omega - is) - s^2 - K_{\ell,s} \Bigg] \hat{R}_s(r) = 0 \ .  \nonumber
\eea 
Note that this radial equation has coefficients that are complex when $s\neq 0$. $K_{\ell,s}$ 
is a separation constant that labels the spheroidal harmonics that can be determined from the angular Teukolsky equation,
\bea\label{eq:spheroidal}
\left[\frac{1}{\sin\theta} \p_\theta \left( \sin\theta \p_\theta \right) + (M\omega\cos\theta-s)^2 - \frac{(m+s\cos\theta)^2}{\sin^2\theta} - s^2 + K_{\ell,s} \right] S_s(\theta) = 0  \ .
\eea
The radial equation \eqref{eq:radial} corresponds to a doubly confluent Heun equation, where both the horizon and infinity ($r=\infty$) are irregular singular points of rank one. In contrast, the angular equation \eqref{eq:spheroidal} is a singly confluent Heun type, with regular singular points at \( \theta = 0 \) and \( \theta = \pi \), while \( \theta = \infty \) is an irregular singular point, also of rank one.
The solutions to the spheroidal equation for a Kerr black hole with arbitrary frequency are challenging to obtain, requiring numerical or approximate methods. In the slowly rotating, low-frequency limit \(a\omega \ll 1\), the separation constant \(K_{\ell, s}\) can be expressed as a power series. For the following sections, we provide the first few terms of this expansion.
\bea\label{separationconst}
K_{\ell,s}=(\ell-s)(\ell+s+1)+s- M\omega \,\frac{2 m (s^2+\ell (\ell+1))}{\ell(\ell+1)}+ \mathcal{O}(a^2\omega^2)\,.
\eea
The parameters satisfy the usual relations for defining angular momenta: 
\(\ell \geq |s|\), and \(-\ell \leq m \leq \ell\), 
where \(\ell\) and \(s\) can be either integers or half-integers, and \(m\) takes all integer values.

The boundary conditions for tidal Love numbers in the context of black holes play a critical role in determining the uniqueness of the solution to the perturbation equations governing tidal deformations. In the case of static (time-independent) tidal Love numbers, the condition of smoothness at the event horizon becomes essential. Regularity ensures that there are no singularities or unphysical behavior near the horizon, which would correspond to perturbations that are not physically reasonable. For dynamical Love numbers, the boundary conditions typically involve requiring ingoing waves at the horizon, reflecting the fact that perturbations should be absorbed by the black hole. 

In the case of extremal Kerr black holes, purely ingoing modes for perturbations with $\omega \ne 0$  take the form
\bea
\hat{R}(r) \propto e^{+\frac{iM(2M\omega - m)}{( r - M)}} (r - M)^{-2 i M\omega}\qquad
{\rm as} \qquad  r \rightarrow M\,,
\eea
as derived from an analysis of the radial Teukolsky equation. These boundary conditions reflect the absorption of perturbations by the black hole at the horizon. For quasi-normal modes (QNMs), outgoing conditions at infinity are also imposed to ensure radiative behavior, ensuring that the perturbations describe waves propagating outward. However, for gravitational tidal deformation coefficients, the boundary conditions at infinity are not fixed, as they correspond to matching the perturbation solutions with the external gravitational field. At \textit{large distances} from the black hole (as \( r \to \infty \)), the perturbation is typically written as a superposition of two terms: the external tidal field and the black hole's response. The first term \( A_l \, r^l \) represents the externally applied tidal field, while the second term \( B_l \, r^{-(l+1)} \) describes the induced response or deformation of the black hole. The ratio of these terms is used to calculate the Love number.

\section{Static Love numbers for Extremal Kerr}
\label{sec:StaticEK}

Now we will look for the static Love numbers for extremal Kerr.
The angular part of the Teukolsky equation \eqref{eq:spheroidal} given that $\omega=0$ reduces to 
\bea
\left[\frac{1}{\sin\theta} \p_\theta \left( \sin\theta \p_\theta \right)  - \frac{(m+s\cos\theta)^2}{\sin^2\theta}  + K_{\ell,s} \right] S_s(\theta) = 0  \ .
\eea
In this case, the equation becomes analytically solvable, enabling us to determine the exact eigenvalues, expressed as follows
\bea
K_{\ell,s}=(\ell-s)(\ell+s+1)+s
\eea
The radial equation \eqref{eq:radial} in the static case becomes 
\bea \label{eq:radialstatic}
&\!\!\!\!\!\!\!\!\!\!&\Bigg[\p_r\left( (r-M)^2 \p_r \right)+  \frac{m\,M\,(m\, M+ 2 i (r-M) \, s)}{(r-M)^2} - s^2 - K_{\ell,s} \Bigg] \hat{R}_s(r) = 0 \ .  
\eea 

It is now valuable to explore different scenarios where the radial equation is solvable. For $m=0$, additional simplifications occur:
\bea
&\!\!\!\!\!\!\!\!\!\!&\Bigg[\p_r\left( (r-M)^2 \p_r \right) - s^2 - K_{\ell,s} \Bigg] \hat{R}_s(r) = 0 \ .  
\eea 
One can then easily find the exact solutions of the radial equation to be
\bea
\hat{R}_s(r)=c_1 \, (r-M)^{\left(-1+\sqrt{(1+2\ell)^2-4 s}\right)/2}+c_2\, (r-M)^{\left(-1-\sqrt{(1+2 \ell)^2-4 s}\right)/2}\,.
\eea
The behavior of the solutions $r\rightarrow \infty$ (and $s=0$) then
\bea
\hat{R}_{s=0}(r) \rightarrow c_1 \, r^{\ell}+c_2\,  r^{-\ell -1}\,.
\eea
which in turn helps us identify the tidal coefficient as the ratio of the amplitudes
\bea
k_{\ell,m=0}= \frac{c_2}{c_1} 
\eea
We proceed to examine the solution at the horizon \(r = M\). To avoid a solution that diverges as \(r \to M\), we must set \(c_2 = 0\). Consequently, all tidal coefficients vanish,
\bea
k_{\ell,m=0}= 0.
\eea
This result is consistent with the vanishing of Love numbers in non-extremal Kerr black holes. It is also found that the Love number vanishes in the more general case where  $s\ne 0$. \\

We now turn to the more general case where \(m \neq 0\) and \(s = 0\). To proceed, we apply the following coordinate transformation, which will allow us to explore the behavior of the equation in this broader context and analyze how the presence of a nonzero azimuthal number \(m\) affects the solution. Taking
\bea\label{eq:coords}
z=\frac{m M }{(r-M)}\,,
\eea
We can show that, by applying the appropriate coordinate transformation, \eqref{eq:radialstatic} can be brought into a form that can be reduced to the spherical Bessel equation. This allows us to simplify the analysis and obtain solutions in terms of known special functions.
\bea\label{eqdifbessel}
z^2 \, \frac{d^2 \hat{R}_{s=0}}{dz^2}+(z^2- \ell(\ell+1)) \hat{R}_{s=0}=0. \label{eq:sphericalbessel}
\eea
The equation has a regular singularity at \(z = 0\) and an irregular singularity of rank 1 at \(z = \infty\). The two solutions to this second order differential equation are related to the spherical Bessel functions of the first and second kinds, given by \(z j_l(z)\) and \(z y_l(z)\).
Furthermore, to impose ingoing boundary conditions at the horizon, it is more convenient to utilize the spherical Hankel functions, which are combinations of \(j_{\ell}(z)\) and \(y_{\ell}(z)\). We consider
\bea
h^{(1)}_{\ell}(z)&=& \left(j_{\ell}(z)+i \, y_{\ell}(z)\right)\,\\
h^{(2)}_{\ell}(z)&=&\left(j_{\ell}(z)-i \, y_{\ell}(z)\right)\,.
\eea
In many contexts, Hankel functions help transition between the behavior near a source (described by Bessel functions) and the asymptotic far-field behavior (described by spherical waves or decaying/radiating fields). The radial solution to \eqref{eqdifbessel} in this basis reads
\bea\label{eqsol}
\hat{R}_{s=0}=\,z (c_1\, h^{(1)}_{\ell}(z)+c_2 \, h^{(2)}_{\ell}(z))\,.
\eea
For large $z$, the asymptotic expressions for the Hankel functions are given by:
\be h_\ell^{(1)} \sim  -i \, e^{i(z-\frac{1}{2}n\pi)} {\rm \, and}  \,\, h_\ell^{(2)}  \sim  i \, e^{-i(z-\frac{1}{2}n\pi)}.\ee
Thus, near the  horizon $r\rightarrow M$ we obtain the following behavior
\bea
\hat{R}_{s=0} \rightarrow \frac{1}{r-M} \  \left[ c_1 \, e^{\frac{im M }{(r-M)}} \, (1+...)+c_2 \, e^{-\frac{im M }{(r-M)}}\,(1+...) \right]
\eea
Now we need to impose smoothness as $r \rightarrow M$.  In extreme Kerr, as $r \rightarrow M$ at finite $t$, 
one reaches an internal infinity and not the event horizon. To reach the future horizon 
one needs to send $r \rightarrow M$ at some fixed instant of retarded time. Were one to take the limit $r \rightarrow M$
at fixed advanced time, one would end up on the past horizon which is an unphysical part of the exact Kerr
solution if one were considering a Kerr black hole being formed by gravitational collapse. In practice, no
perturbation will ever have exactly zero frequency. If one looked at time dependent perturbations of Kerr on the future 
horizon, and then took the limit as $\omega \rightarrow 0$, then one would set $c_1=0$.

The usual procedure involves analytically continuing the functions into the asymptotic region $r\rightarrow \infty$. The behavior of the solutions at large $r$ yields
\bea\label{eq:lovedef}
\hat{R}_{s=0} \rightarrow \tilde{c}_2 \,  r^{\ell}\left(1+ k_{\ell m}\, \left(\frac{r}{2M}\right)^{-(1+2\ell)}+...\right)\,.
\eea
From this expression, the tidal coefficient reduces to \footnote{Note that here we can also consider $\Gamma(\ell+1/2)= (2\ell)! \sqrt{\pi} / (\ell! 4^{\ell} )$.}
\bea\label{eq:lovescalar}
k_{\ell m}=- \frac{ 2 \pi i \,m^{2 \ell+1}}{2^{2( 2 \ell+1)}\Gamma\left(\frac{1}{2} + \ell\right) \Gamma\left(\frac{3}{2} + \ell\right)}\,.
\eea
In terms of dissipation, it is important to note that a larger tidal Love number typically signifies greater dissipative behavior. For extremal Kerr black holes, these dissipative losses are generally much smaller -- reaching their minimum -- compared to non-extremal Kerr black holes, though they remain present in these rotating systems. As a result, extremal black holes dissipate significantly less energy than their non-extremal counterparts, but they are not entirely lossless. Drawing a comparison with superconductors can provide valuable insights. Superconductors can also dissipate energy through the dissipation of magnetic energy when subjected to changing magnetic fields. However, a perfectly ideal superconductor would not dissipate energy when rotating in a constant magnetic field. Extremal Kerr black holes exhibit dissipation, deviating from the behavior expected of a superconductor.
 
For stars with positive tidal Love numbers, dissipation often occurs through mechanisms such as viscous heating, where the internal friction converts the kinetic energy of tidal motions into heat. On the other hand, stars with negative tidal Love numbers can experience dissipation through processes such as resonant tidal excitation of internal waves. Given that the tidal dissipative coefficient is negative, we argue that both extremal and non-extremal Kerr black holes will posses internal dissipative processes much like superconductors.\\

Lastly, we look at the case with arbitrary spin parameter $s\ne0$. Employing the transformation \eqref{eq:coords}, the radial equation becomes
\bea
z^2 \, \frac{d^2 \hat{R}_{s}}{dz^2}+(z^2+ 2 i  z s - \ell(\ell+1)) \hat{R}_{s}=0
\eea
This is Whittaker's equation, which is closely related to the confluent hypergeometric equation. It has a regular singularity at the origin with indices, and an irregular singularity at infinity of rank one. The solutions are
\bea
 \hat{R}_{s}= c_1 \,M_{s,\ \ell+{1/2}}\,(2i z) +c_2 \, W_{s, \ \ell+{1/2}}\,(2i z)\,.
\eea
where taking $\bar{\kappa}=  \ell+{1/2}$ the two Whittaker functions $M_{s,\bar{\kappa}}(x),W_{s,\bar{\kappa}}(x)$ are related to the  confluent hypergeometric functions by
\bea M_{s,\bar{\kappa}} = e^{-z/2}\, z^{\bar{\kappa}+{1/2}}\,M(\bar{\kappa} -s +{1/2},1+2\bar{\kappa},z) \eea
and
\bea W_{s,\bar{\kappa}} = e^{-z/2}\, z^{\bar{\kappa}+{1/2}}\, U(\bar{\kappa} -s +{1/2},1+2\bar{\kappa},z). \eea

Regularity at the horizon at $z=\infty$, fixes $c_1=0$. By plugging in the explicit expression of $z$ in \eqref{eq:coords} we can express the solution as a function of the radial coordinate $r$. Taking the following asymptotic expansion at large distances, the smooth wave solution as $r\rightarrow \infty$, is given by
\bea
 \hat{R}_{s} \rightarrow c_2 \left[r^{\ell} (m M)^{-\ell} \left(\frac{(2 i)^{-\ell} \Gamma (2 \ell+1)}{\Gamma (\ell-s+1)}+...\right)+ r^{-\ell-1} (m M)^{\ell+1}  \left(\frac{(2 i)^{\ell+1} \Gamma (-2 \ell-1)}{\Gamma (-\ell-s)}+...\right)\right]
\eea
where dots denote the polynomial corrections of the form $\frac{1}{r^n}$ with  $n>0$, a positive integer. 
By comparing with equation \eqref{eq:lovedef} we observe that the static tidal deformation coefficient for an extremal Kerr black hole is given by
\bea
k_{\ell m}=- \frac{( i m)^{2 \ell+1}\Gamma\left(-1-2 \ell\right) \Gamma\left(1+ \ell - s \right)}{\Gamma\left(1+ 2 \ell\right) \Gamma\left(- \ell- s\right)}\,.
\eea
As a consistency check, it is important to note that for the scalar case $s=0$, we recover the expression we previously derived in equation \eqref{eq:lovescalar}.
We conclude this section by stating that for extremal Kerr black holes, the static Love numbers are either finite or zero, with no indication of divergence or singular behavior.

\section{\bf Dynamical Love numbers for Extremal Kerr Black Holes}
\label{sec:DynEK}

In this section we find an expression for the dynamical Love number of extremal Kerr black holes using the Leaver-MST approach. For simplicity we will consider black holes with equal momentum parameter $a$ and mass $M$ with $a=M=r_+=1/2$.

\subsection{Via MST-Leaver Method}

The disagreements over whether the Love number vanishes or not have now been resolved. The disagreement sparked a debate on the gauge invariance of the definition of tidal Love numbers. In cases like rotating Kerr black holes, where the perturbation equation is only exactly soluble in terms of doubly confluent Heun functions, different near-region approximations may be used to determine the exact form of these coefficients \cite{Hui:2022vbh,Charalambous:2021kcz,Perry:2023wmm}. However, this has led to a new debate about the ambiguity found in various near-zone geometry prescriptions \cite{Lowe:2011aa,Cvetic:2024dvn}. In this section, we present a gauge-invariant definition for the extremal Kerr black hole tidal coefficient that does not rely on any near-zone approximation. Our novel approach employs the MST-Leaver method, as described below.

The MST-Leaver method is an analytical technique for solving differential equations using series expansion of solutions. Leaver obtained solutions to the Teukolsky equation expressed as a series of Coulomb wave functions \cite{Leaver:1986vnb}. This was later reformulated by MST \cite{Mano:1996vt}. A key point in the MST construction of solutions has been providing series representations for all radial solutions in terms of the same series coefficients (namely, $a_n$ as defined later in this section). Since the differential equations involved do not permit globally convergent solutions, the method relies on patching solutions with different overlapping regions of convergence. While having a single series expansion would be more convenient, in certain cases, such as the doubly confluent Heun equations for extremal Kerr black holes, two different sets of series expansions are necessary, each valid in a specific region. Our choices for the radial solutions, which are based in Eqs. (60a) - (61b) in Ref. \cite{Casals:2018eev} are
\bea\label{solutions}
&& R_H^{\nu} \qquad \text{and} \qquad R_H^{-\nu-1} \qquad  \text{for} \qquad r_+\le r < \infty \,,\\
&& R_C^{\nu} \qquad \text{and} \qquad R_C^{-\nu-1} \qquad  \text{for} \qquad r_+< r \le \infty
\eea
where the subscript $H$ refers to solutions valid on the horizon and those at the asymptotic boundary with $R^{\nu}_C$ referring to Coulomb wavefunctions \footnote{We note that we choose to use the subindex $C$ for the solutions, which correspond to $R^{\nu}_{\infty,+}$ in Eq. (60.a) in \cite{Casals:2018eev} and with $R^{\nu}_H$  correspond to those with $R^{\nu}_{0,+}$  in Eq. (61.a) in \cite{Casals:2018eev} }. The auxiliary parameter $\nu$, the so-called renormalized angular momentum, is introduced as a generalization of the eigenvalue $\ell$ in \eqref{separationconst}, taking on non-integer values. This modification ensures that the Teukolsky equation takes a form where the solution is explicitly invariant under the transformation $\nu \rightarrow -\nu - 1$, allowing it to be solved using an expansion in terms of hypergeometric functions. For more details, refer to \cite{Casals:2018eev}.

In physical applications, it is natural to consider the solutions to the radial equation \eqref{eq:radial} according to the boundary conditions imposed at the horizon $r=r_+$ and at the boundary $r=\infty$. In the extremal Kerr problem, based on purely ingoing radiation entering the black hole, the solutions satisfy the following boundary conditions
\bea
&& \hat{R}_s(r)\equiv  \hat{R}^{\text{in}}_s(r)  =
\left\{
\begin{aligned}\label{eq:TRcoeffs}
 \mathcal{T}_{in} e^{\frac{i M (2 M \omega - m)}{r-M}} (r-M)^{-2 i M\omega -2 s} \qquad \text{when} \,\, r\rightarrow M\,\\
\mathcal{I}_{in} \,r^{-2i M \omega-1} \,e^{-i\omega r}+ \mathcal{R}_{in}\, r^{2i M \omega-1-2s} \,e^{i\omega r} \,  \, \text{when } \,\, r \to \infty
\end{aligned}
\right.
\eea
where $ \mathcal{I}_{in} ,\mathcal{T}_{in}, \mathcal{R}_{in}$ are respectively, complex valued incident, transmission and reflection coefficients of the solutions.

After matching \eqref{solutions} within the large overlap region of convergence \( r_+ < r < \infty \), the radial solution \( R_H \) that is ingoing at the event horizon can be expressed in the Coulomb basis \( R_C \), making it manifestly invariant under the transformation \(\nu \rightarrow -\nu - 1\) and also valid asymptotically at \( r = \infty \), as follows \footnote{See Eqn. (71) in \cite{Casals:2018eev} for more details }:
\bea\label{base0}
 \hat{R}_s(r) = K_{\nu} \,R^{\nu}_C + K_{-\nu-1} \, R_C^{-\nu-1}\,, \qquad  \text{for} \qquad r_+< r \le \infty \,.
\eea
This series representation of the ingoing solution on the black hole event horizon is valid at $r=\infty$ and will be used in this section to find the Love numbers.

The functions used here are defined by
\bea\label{eq:coefs}
R^{\nu}_C&=&\zeta^{(\infty)}_{+} e^{i k/(2x)} e^{i\omega x} x^{-s+\nu} (2\omega)^{\nu+1}e^{-i\pi\chi_{s}/2}e^{-i\pi(\nu+(1/2))}\\
&\times& \sum_{n=-\infty}^{\infty}a_{n}^\nu\,(-2i\omega x)^{n}\, \frac{\Gamma(q_n^{\nu}+\chi_s)\Gamma(1-2q_n^{\nu})}{\Gamma(q_n^{\nu}-\chi_{s})\Gamma{(1-q_n^{\nu}+\chi_s)}} M(q_n^{\nu}+\chi_{s},2 q_n^{\nu},-2i\omega x)\,,\nonumber\\
R^{-\nu-1}_C&=&\zeta^{(\infty)}_{+} e^{i k/(2x)} e^{i\omega x}(2\omega)^{\nu+1} x^{-s-\nu-1} (-2i\omega)^{-2\nu-1}e^{-i\pi\chi_{s}/2}e^{-i\pi(\nu+(1/2))}\\
&\times& \sum_{n=-\infty}^{\infty}a_{n}^{\nu}\,(-2i\omega x)^{-n}\, \frac{\Gamma(2q_n^{\nu}-1)}{\Gamma(q_n^{\nu}-\chi_{s})} M(1-q_n^{\nu}+\chi_{s},2 (1-q_n^{\nu}),-2i\omega x) \,. \nonumber
\eea
 in terms of regular confluent hypergeometric functions $M(a,b,y)={}_1 F_1(a,b;y)$. Here instead of the radial coordinate $r$ we shall use the shifted radial coordinate
\bea
x=r-r_+ \,.
\eea
We  define shifted frequency parameter 
\bea
k=(\omega- m)\,
\eea
as well as
\bea
\chi_{s}=s-i\omega, \qquad  q_n^{\nu}=n+\nu+1
\eea

The $a^{\nu}_{n}$ are series coefficients that satisfy the following recurrence relation
\bea
\alpha_n a^{\nu}_{ n+1} + \beta_n a^{\nu}_{ n} + \gamma_n a^{\nu}_{ n-1} = 0\,, \qquad n \in \mathbb{Z}
\eea
with coefficients given by
\bea\label{eq:recursive}
\alpha_n &=& \frac{\epsilon(q^{\nu}_n + \chi_s)(q^{\nu}_n - \chi_{-s})}{q^{\nu}_n(2q^{\nu}_n + 1)}\,,\\
\beta_n &=& (q^{\nu}_n - 1)q^{\nu}_n - _s\bar{A}_{\ell m \omega} - \epsilon \frac{\chi_s \chi_{-s}}{(q^{\nu}_n - 1)q^{\nu}_n}\\
\gamma_n &=& \frac{\epsilon(q^{\nu}_n - 1 - \chi_s)(q^{\nu}_n - 1 + \chi_{-s})}{(q^{\nu}_n - 1)(2q^{\nu}_n - 3)}
\eea
 where we have defined the quantities and$ _s\bar{A}_{\ell m \omega }= -\frac{7}{4}\omega^2 + s(s + 1) + K_{\ell,s}-s$ 
\bea
\epsilon= \omega k\,.
\eea 
For the series coefficients, we adopt the normalizations $a_0^{\nu}=a_0^{-\nu-1}=1$, which consequently imply $a_{-n}^{\nu}=a_{n}^{-\nu-1}$ for all integer values of $n$.

We can interpret the coefficients $K_{\nu}, K_{-\nu-1}$ in the solution \eqref{base0} as connection coefficients with the boundary functions that are manifestly symmetric under $\nu \rightarrow -\nu - 1$. Consequently, one can explicitly define the expression with $\nu$ indices
\bea\label{eq:Kpar}
K_{\nu} = \frac{\zeta^{(0)}_+}{\zeta^{(\infty)}_+} k^{\nu+1} (-ik)^{-2\nu-1} (2\omega)^{-\nu-1} e^{i\pi s}\frac{ \sum_{n=p}^{\infty} C^{\nu}_{n,n-p}}{ \sum_{n=-\infty}^{p} D^{\nu}_{n,p-n}}
\eea
where
\bea
C^{\nu}_{n,j}=\frac{\Gamma(q^{\nu}_n + \chi_s) \Gamma(2q^{\nu}_n - 1) (1 - q^{\nu}_n + \chi_{-s})_{j} }{ \Gamma(q^{\nu}_n - \chi_s) \Gamma(q^{\nu}_n + \chi_{-s})(2 - 2q^{\nu}_n)_{j} \,j!} (-ik)^{j-n}}{a^{\nu}_n\\
D^{\nu}_{n,j}=\frac{\Gamma(q^{\nu}_n + \chi_s) \Gamma(1-2q^{\nu}_n ) ( q^{\nu}_n + \chi_{-s})_{j} }{ \Gamma(q^{\nu}_n - \chi_s) \Gamma(1-q^{\nu}_n + \chi_{s})( 2q^{\nu}_n)_{j} \,j!}(-2i \omega)^{n+j}}{a^{\nu}_n \,.
\eea 
Here, $p$ represents an arbitrary integer, which, without loss of generality, will later be set to $p = 0$. The term $(z)_n = \frac{\Gamma(z+n)}{\Gamma(z)}$ denotes the Pochhammer symbol, where $\Gamma$ is the Gamma function.
The normalization constants $\zeta^{(0)}_{+} $ and $\zeta^{(\infty)}_{+}$ were specified to ensure radial solutions exhibit symmetry under $\nu \rightarrow -1 - \nu $. The explicit expressions  are
\bea
\zeta^{(0)}_{+} = k^{-\nu} (-ik)^{\nu} e^{i\pi \nu} \left( \frac{\sin(\pi(\nu - i\omega))}{\sin(\pi(\nu + i\omega))} \right)^{1/2}\\
\zeta^{(\infty)}_{+}= \omega^{-\nu} (-i\omega)^{\nu} e^{i\pi \nu} \left( \frac{\sin(\pi(\nu - i\omega))}{\sin(\pi(\nu + i\omega))} \right)^{1/2}
\eea
{\bf Low frequency expansion}\\
So far, we have considered exact solutions of the Teukolsky equation. Now, let us consider their low frequency approximations and determine the value of $\nu$. Explicitly, the renormalized angular momentum coefficient is determined through iterative solutions of the equation to higher orders. For instance, when $|\omega|$ is sufficiently small we get
\bea\label{integerpar}
\nu \equiv \ell +\Delta \ell + \mathcal{O}(\omega^3)
\eea
where
\bea
\Delta \ell  =\frac{-15\ell^{4} - 30\ell^{3} - 6\ell^{2}s^{2} - 4\ell^{2} - 6\ell \, s^{2} + 11\ell - 3\,s^{4} + 6\,s^{2}}{2\ell(\ell + 1)(2\ell + 1)(4\ell^{2} + 4\ell - 3)} \,\, \omega^2\,.
\eea
where $\ell$ is an integer number satisfying $\ell \ge |s|$. Higher order corrections can be found in equation (94a) and (94b) in \cite{Casals:2018eev}.  

For practical reasons, and to ensure a reliable, gauge-invariant analytical definition of the tidal coefficients, it is preferable to use the asymptotic coefficients \( K_{\nu, -\nu-1} \) defined in \eqref{eq:Kpar}, rather than those defined in the near-horizon region \( r \rightarrow r_+ \). As we will demonstrate, the renormalized symmetry provides a prescription for determining the Love numbers in the low-frequency regime \( \omega \ll 1 \). We will also argue that this low-frequency regime can be effectively considered analogous to the region near the black hole.

\subsection{Love numbers in the low frequency regime}

Having previously described the solutions to the radial equation \eqref{base0} manifestly invariant under $\nu\rightarrow -1-\nu$, we can now derive explicit expressions for the radial solution in the small-$\omega$ limit and identify the tidal Love coefficients. 

To proceed, we first examine the expressions for the coefficient \(a_n^{\nu}\) appearing in $R^{\nu}_C$ in the small \(\epsilon\) regime. Focusing on the leading terms in the small-\(\epsilon\) expansion, it was shown in \cite{Casals:2018eev} that
\bea
a_n^{\nu}= \tilde{a}^{\nu}_n  \,\epsilon^{|n|} + \mathcal{O}(\epsilon^{|n|+1})  \qquad \text{for} \qquad n \in Z
\eea
Explicit expression for the $a_n^{\nu}$ coefficients can be obtained numerically or prescribed analytically in certain regimes, in an $\epsilon$ expansion. To this end one first defines the ratios
\bea
R_n=\frac{a_n^{\nu}}{a_{n-1}^{\nu}}\,\qquad L_n=\frac{a_n^{\nu}}{a_{n+1}^{\nu}}
\eea
such that expressing the ratios as continued fractions one obtains
\bea
a_{n}^{\nu}=-\frac{\gamma_n}{\beta_n+\alpha_n R_{n+1}} \, a_{n-1}^{\nu}\qquad \text{for} \qquad n>0 \,\\
a_{n}^{\nu}=-\frac{\alpha_n}{\beta_n+\gamma_n L_{n-1}} \, a_{n+1}^{\nu}\qquad \text{for} \qquad n<0 
\eea
given a certain normalization choice for the  coefficients $a_0^{\nu}=a_0^{-\nu-1}$ such the ones defiend above.
Taking this in consideration we find the leading terms in the coefficients, which satisfy
\bea\label{avalues}
a_1^{\nu}=a_{-1}^{-\nu-1}
\sim \frac{(1-s+\ell)^2}{2 (1+\ell)^2(1+2\ell)}\epsilon+\mathcal{O}(\epsilon^2) \equiv \tilde{a}_{1}^{\nu} \epsilon +\mathcal{O}(\epsilon^2) \\
a_{-1}^{\nu}=a_{1}^{-\nu-1} \sim - \frac{(s+\ell)^2}{2 \ell^2(1+2\ell)}\epsilon+\mathcal{O}(\epsilon^2)\equiv \tilde{a}_{-1}^{\nu} \epsilon +\mathcal{O}(\epsilon^2)
\eea

We have explicitly verified that the bilateral recurrence relation defined in \eqref{eq:recursive} aligns with the expressions given in equation (2.8) of \cite{Mano:1996vt} for extremal black holes, under the parameter limits $\kappa \to 0$ and $q \to 1$ specified therein.

From \eqref{eq:coefs} we readily obtain\footnote{To systematically identify the growing and decaying terms in the radial direction \(r\), we first analyze the expansion of \(R^{\nu}_C\)
in the small \(\omega\) limit. Notably, in the series expansion of \(R^{\nu}_C\), only terms with \(n \leq 0\) contribute. Conversely, in \(R^{-1-\nu}_C\), the series expansion involves contributions solely from terms with \(n \geq 0\).}
\bea\label{base1}
 \hat{R}_s(r) &=&   \, K_{\nu}  \,e^{i k/(2r)} e^{i\omega r} e^{-i\pi\chi_{s}/2} \,r^{\nu-s}\, \left( F^0_{\nu}+ \frac{F^1_{\nu}}{r}+... \right)  \\
 &+&  \, K_{-\nu-1}\, \,e^{i k/(2r)} e^{i\omega r} e^{-i\pi\chi_{s}/2}\,r^{-\nu-1-s}\,  \left( F^0_{-\nu-1}+\frac{ F^1_{-\nu-1}}{r}+... \right)\,, \,  \text{for} \,\, \omega \ll 1
 \eea
where the coefficients are defined as
\bea \label{eq:Feq}
F^j_{\nu}&=& \omega ^{\nu +1}   (-2 i)^{\nu -j+1}(-m)^{j}  (f^{(0)}_{\nu,j} + f^{(1)}_{\nu,j} \, \omega +\mathcal{O}(\omega^2))\\
F^j_{-\nu-1}&=& F^j_{\nu\rightarrow -\nu-1} \label{eq:Feq11}
\eea
for all values of the spin parameter $a$ where e.g. for $j=0$ we find
\bea
 f^{(0)}_{\nu,0}&=&-(\pi)^{-1} \sin (\pi  (\nu-s)) \, \Gamma (-2 \nu -1) \, \Gamma (\nu +1+s) \,,\\
 f^{(1)}_{\nu,0}&=&  f^{(0)}_{\nu,0}  \left(m \, (2 \nu +1)\, \tilde{a}_{-1}^{\nu}- i \,\psi (s+\nu +1)\right)
 \eea
 where $\psi$ is the polygamma function.

Note that the relevant terms in the slow frequency expansions \eqref{eq:coefs} are those with $n \le 0$. We have also further considered the asymptotic behavior $(r-r_+) \rightarrow r$. 
The response coefficients can be easily identified in the low-frequency regime $\omega \ll 1$ from
 \bea\label{base3}
 \hat{R}^{\text{in}}_s(r) =  \,r^{\nu-s} c_0 \left[ \,\left(1+\frac{\tilde{F}^1_{\nu}}{r}+... \right) + r^{-2\nu-1-2s}  \frac{F^0_{-\nu-1} \, K_{-\nu-1}}{F^0_{\nu} \, K_{\nu}}\, \left(1+\frac{\tilde{F}^1_{-\nu-1}}{r}+... \right) \right]\,, 
 \eea
where we defined the factor $c_0=F^0_{\nu} \, K_{\nu}  \,e^{i k/(2r)} e^{i\omega r} e^{-i\pi\chi_{s}/2}$. The tidal Love number is defined as the coefficient of the asymptotic terms of the form $r^{-2\ell-1-2s}$ with $\ell \in N$, which characterize the object's response to an external tidal field in the low-frequency limit. 
\bea\label{eq:Lovepartial}
k_{\ell m} =  \left. \lim_{\Delta \ell \rightarrow 0}  (1- 2 \Delta \ell \log(r))\, \frac{ F^0_{-\nu-1}}{F^0_{\nu}} \frac{K_{-\nu-1}}{K_{\nu} } \right|_{\nu =\ell+\Delta \ell }\,,
\eea
where the logarithmic terms come from the expansion
\bea
r^{-2\nu-1} = r^{-2\ell-1}(1-2 \Delta \ell \log(r) +\mathcal{O}( \Delta \ell ^2))
\eea

Alternative definitions of the Love number,  differing from those proposed here which consider only the ratio $K_{-\nu-1}/K_{\nu}$ are discussed in \cite{Bautista:2023sdf}.

Taking all the previous results into consideration, we can analyze the tidal Love number \eqref{eq:Lovepartial} in the the low frequency regime  $\omega\ll 1$. Considering \eqref{eq:Kpar} we find that
\bea
\frac{K_{-\nu-1}}{K_{\nu}}&=&(2 m \omega)^{2\nu+1}\frac{\Gamma(-1-2\nu)^2\Gamma(1-s+\nu)\Gamma(1+s+\nu)}{\Gamma(-s-\nu)\Gamma(s-\nu)\Gamma(1+2\nu)^2} \\
&&\times  \left[1+\left( 2m(1+2 \nu)P+ 2\pi i \cot(\pi\nu))\right) \,\omega+\mathcal{O}(\omega^2)\right] \nonumber
\eea 
where we defined $P= \frac{(s^2+\nu^2)}{(s+\nu)^2}\tilde{a}^{\nu}_{-1}  + \frac{(s^2+(1+\nu)^2)}{(1-s+\nu)^2}\tilde{a}^{\nu}_{1}$. The ratio among the coefficients
\bea
\frac{F^0_{-\nu-1}}{F^0_{\nu}}&=&-(2 i \omega)^{-2\nu-1}\frac{\Gamma(1+2\nu)}{\Gamma(-2\nu-1)}\frac{\Gamma(s-\nu)}{\Gamma(1+\nu+s)} \\
&&\times  \left[1-(m \,(1+2 \nu) \left( \tilde{a}^{\nu}_{1}  + \tilde{a}^{\nu}_{-1}) + i\psi(1+s+\nu)-i \psi(s-\nu)\right) \,\omega+\mathcal{O}(\omega^2)\right] \nonumber
\eea 
defined by assessing the limit in equation \eqref{eq:Feq} and \eqref{eq:Feq11}. 

 A straightforward computation,  replacing these results in equations \eqref{eq:Lovepartial} and \eqref{avalues}, yields the frequency-dependent tidal response coefficient 
 \bea\label{eq:Love}
k_{\ell m}=-  \left. \lim_{\Delta \ell \rightarrow 0} (1- 2 \Delta \ell \log(r))\,   \,( -i m)^{2 \nu+1}\frac{\Gamma(-2\nu-1)\Gamma(\nu-s+1)}{\Gamma(2\nu+1)\Gamma(-\nu-s)}\left(1+ Q_{\nu} \,\omega+\mathcal{O}(\omega^2)\right)  \right|_{\nu =\ell+\Delta \ell }\,,
\eea
where
\bea
 Q_{\nu}=i   \left(\psi(1+s+\nu)-\psi(s-\nu)+2 \pi  \cot (\pi  \nu )\right)-\frac{1+2 \nu }{m}-\frac{m s (1+2 \nu) (s-2 \nu  (\nu +1))}{2 \nu ^2 (\nu +1)^2}
 \eea
 It is also important that we expand the renormalized momentum $\nu=\ell+\Delta\ell$ where $\ell$ is an integer. We obtain
\bea\label{eq:Love}
\boxed{k_{\ell m}=-  \frac{(-1)^{s} \, i \,(m)^{2 \ell+1}\, (\ell-s)!(\ell+s)!}{2(2\ell)!(2\ell+1)!} \left(1+ Q_{\ell} \,\omega+\mathcal{O}(\omega^2)\right) \,,}
\eea
where 
\bea
Q_{\ell}=  i \left(\sum^{2s-1}_{j=0}\frac{1}{\ell+j+1-s}-2  \log(r) \right) -\frac{1+2 \ell }{m}-\frac{m s (1+2 \ell) (s-2 \ell  (\ell +1))}{2 \ell ^2 (\ell +1)^2} \,.
\eea
For $s=0$ the first term that appears in the last equation vanishes.  We use only the finite parts, and the singular contribution can be ignored \footnote{The full solution $ \hat{R}^{\text{in}}_s(r) $ in \eqref{base3} is regular, and he divergent terms get canceled by a similar singularity in the source series.} We find that finite terms come from expansion to linear order in $\Delta \ell$. The limit $\Delta \ell \rightarrow 0$ of Eq.\eqref{eq:Love} is complicated by the presence of poles infinite gamma functions and $\cot$. The presence of the pole also generates a finite logarithmic contribution. We obtain 
\bea
\frac{\Gamma(-2\nu-1)\Gamma(\nu-s+1)}{\Gamma(2\nu+1)\Gamma(-\nu-s)}=  \frac{(-1)^{\ell+s+1}(\ell-s)!(\ell+s)!}{2(2\ell)!(2\ell+1)!} + \mathcal{O}(\Delta \ell ^0)
\eea

 along with
 \bea
 \psi(1+s+\nu)-\psi(s-\nu)=
 \left\{
\begin{aligned}
&- \frac{1}{\Delta \ell} + \sum^{2s-1}_{j=0}\frac{1}{\ell+j+1-s} + \mathcal{O}(\Delta \ell ^1) \qquad \text{for} \qquad s\ne0 \\
&- \frac{1}{\Delta \ell}+ \mathcal{O}(\Delta \ell ^1)   \qquad \text{for} \qquad  s=0
\end{aligned}
\right.
 \eea

The gravitational tidal coefficients \( k_{\ell m} \), defined precisely in \eqref{eq:Love} at a given frequency \( \omega \) are generally complex and are conveniently expressed as the sum of their real and imaginary parts:

\[
k_{\ell m}(\omega) = \kappa_{\ell m}(\omega) + i \nu_{\ell m}(\omega),
\]

where \( \ell \) and \( m \) denote the conventional spherical harmonic indices. The conservative response, defined by the real part, \( \kappa_{\ell m} \), encodes the tidal deformation Love numbers, while the imaginary part, \( \nu_{\ell m} \), represents dissipative effects. This distinction has a clear physical analogy in electromagnetism: the real part corresponds to conservative effects like refraction, whereas the imaginary part reflects dissipation, akin to the imaginary component of the electric susceptibility. This allows us to define the Love numbers order by order

\[
k_{\ell m}(\omega) = \sum^{\infty}_{n=0} ( \kappa^{(n)} + i \, \nu^{(n)} ) \, \omega^n,
\]
where
\bea
&&\kappa^{(0)}=0\,,\\
&&\nu^{(0)}=-  \frac{(-1)^{s} \,(m)^{2 \ell+1}\, (\ell-s)!(\ell+s)!}{2(2\ell)!(2\ell+1)!}\, \\
&&\kappa^{(1)}=-\nu^{(0)} \left( \sum^{2s-1}_{j=0}\frac{1}{\ell+j+1-s}+ 2  \log(r) \right) \, \\
&&\nu^{(1)}=- \nu^{(0)}  \left( \frac{1+2 \ell }{m}+\frac{m s (1+2 \ell) (s-2 \ell  (\ell +1))}{2 \ell ^2 (\ell +1)^2}  \right)\,
\eea

In this context, we observe three key aspects. First, the dynamical Love numbers for extremal Kerr black holes exhibit both real and imaginary components, with all contributions proportional to the dissipative zeroth-order case. The proportionality of the coefficients for non-extremal Kerr black holes was previously noted in \cite{Perry:2023wmm}. Second, the tidal response coefficients do not vanish for extremal black holes, even beyond the static regime. Our results indicate that extremal black holes, despite having zero temperature, do dissipate energy. This finding is reflected in the non-vanishing tidal dissipative coefficients, both in the static and dynamical regimes, highlighting that extremality does not eliminate dissipation in such systems.

Third, to address uncertainties in the tidal coefficients, particularly those arising from extending Newtonian results to a general relativistic framework, analytic continuation methods have been employed. This technique, pioneered by Kol and Smolkin \cite{kol2012specul}, involves treating the parameter \( \ell \) as a non-integer value to explore higher-dimensional generalizations of GR solutions. Alternatively, the point-particle effective field theory (EFT) framework for binary inspirals provides another method to address these uncertainties, offering a systematic and consistent approach to gravitational dynamics. As an independent check in the scalar case when \( s = 0 \), these results can also be derived using an alternative approach based on perturbation theory and near-zone approximations and EFT matching (see \cite{Austin2024} for details).

Lastly, it is worth noting that the expression for the dynamical Love numbers \eqref{eq:Love} matches the static tidal Love numbers definitions (both dissipative and non-dissipative terms)
\bea
\lim_{\omega\rightarrow 0} \frac{k_{\ell m}}{k^{\omega=0}_{\ell  m}}=1\,.
\eea

In this section, we have focused on analyzing solutions to the wave equation in the extremal Kerr geometry to extract the tidal response. Alternatively, one can reformulate the problem to focus on studying the scattering coefficients \eqref{eq:TRcoeffs}. This approach involves determining the radial coefficients associated with the incident and reflection amplitudes. To obtain the radial coefficients for the incident and reflection amplitudes at radial infinity, one can further decompose \( R_C^{\nu} \) and \( R_C^{-1-\nu} \) in Eq.~\eqref{base0} into two components: one that is purely ingoing and another that is outgoing at infinity. The asymptotic radial solution can thus be expressed using the decomposition in Eq.~\eqref{eq:TRcoeffs}. The coefficients of these ingoing and outgoing components correspond to the transmission and reflection coefficients of the solutions.
Mathematically, this stems from the ability to rewrite the \textit{regular} confluent hypergeometric functions \( M(a, b, y) \) in Eq.~\eqref{eq:coefs} in terms of the \textit{irregular} confluent hypergeometric functions \( U(a, b, y) \). This approach could be advantageous if one aims to compute the scattering amplitude at infinity.
However, we argue that extracting the Love coefficients requires the function to retain a symmetric form under the transformation \( \nu \to -1-\nu \), as is the case with the \( M \)-function. When expressed in terms of the \( U \)-functions, this manifest symmetry \( \nu \to -1-\nu \) is lost, making it impossible to fully determine the tidal deformation coefficients. Thus, while the \( U \)-function decomposition may be useful for scattering calculations, it is less practical for extracting Love coefficients in a symmetric framework.

\section{Discussion}
\label{sec:Disc}

In this work, we have examined the static and dynamic gravitational tidal Love numbers of extremal Kerr black holes, focusing on the unique properties these maximally spinning, zero-temperature black holes exhibit in response to external tidal fields. 

Our findings confirm that extremal Kerr black holes possess finite Love numbers, lending support to recent theoretical work suggesting that black holes, even under extreme conditions, exhibit measurable deformations. By applying the Leaver-MST method, we have introduced a gauge-invariant framework for defining the tidal Love number, thus addressing previous challenges related to ambiguities in defining near-horizon geometry approximations.

In the static limit, from our calculation we observed that extremal Kerr black holes exhibit vanishing Love numbers, consistent with the behavior of their non-extremal counterparts, indicating minimal classical tidal deformability. However, in the dynamical regime, we find that dissipative coefficients remain non-zero, highlighting that these black holes retain some degree of dissipation. This behavior, intriguing in the context of extremal conditions, draws an analogy to superconductors, which also exhibit energy dissipation under varying external fields. However, unlike idealized superconductors, extremal black holes maintain a finite dissipative response even at zero temperature.

In light of recent findings by Horowitz et al. \cite{Horowitz:2023xyl} which assert that higher-curvature corrections lead to curvature singularities and infinite tidal forces at the horizon of extremal Kerr black holes, our results present an alternate perspective by demonstrating that the tidal Love numbers for these black holes remain finite. This discrepancy suggests that extremal Kerr black holes may retain finite tidal responses under gravitational perturbations, contrasting with the picture of horizon singularities and unbounded tidal forces described for theories with higher order corrections employed by Horowitz et al.

Our approach, utilizing the MST-Leaver method, allows us to circumvent near-zone geometry ambiguities by introducing a gauge-invariant treatment of tidal deformations. This approach supports the notion that extremal Kerr black holes possess finite Love numbers even as they approach maximal spin, indicating that the gravitational response to external fields may remain well-defined without invoking horizon singularities. The contrasting results suggest that certain assumptions in the higher-curvature perturbative methods, as applied by Horowitz et al., may introduce divergent behaviors that are not inevitable in other formalisms. Further investigation is warranted to understand the conditions under which extremal Kerr black holes exhibit singular vs. finite tidal responses, as well as to explore how different treatments of the near-horizon geometry might reconcile or deepen these theoretical divergences.

The implications of these findings extend to gravitational wave research, where black hole tidal interactions could meaningfully impact gravitational wave signatures. Given the non-trivial Love numbers of extremal Kerr black holes, these tidal deformations may contribute detectably to the waveforms observed by gravitational wave detectors.

Our analysis can be extended in several directions. First, using the methods developed in this paper, one could perform a systematic computation of Love numbers, including their dependence on higher-order frequency corrections, such as terms proportional to \(\omega^2\). Eventually, it will be important to explore the symmetries and relationships associated with higher-order tidal deformation corrections. Second, looking ahead, this gauge-invariant formalism could be applied to study other compact objects or alternative black hole configurations. Moreover, the renormalized symmetry \(\nu \rightarrow -\nu - 1\) observed in our approach suggests a possible connection to hidden $SL(2,R)$ symmetries, previously proposed to explain near-horizon dynamics of extremal black holes. Exploring this symmetry in depth could yield further insights into the fundamental nature of extremal black hole interactions.

\section*{Acknowledgements} 

We would like to thank Austin Joyce, Luca Santoni, Adam Solomon, Daniel Glazer and Luis Fernando Temoche for discussions. We gratefully thank the Mitchell Family Foundation at Cook's Branch workshop and the Centro de Ciencias de Benasque Pedro Pascual for their warm hospitality. The work of MP is supported by an STFC consolidated grant ST/L000415/1, String Theory, Gauge Theory and Duality. MJR is supported through the NSF grant PHY-2309270.

\bibliographystyle{jhep} 
\bibliography{bibliography} 

\end{document}